\documentclass[12pt]{article}
\pdfoutput=1

\usepackage{amsmath}
\usepackage{amsfonts}
\usepackage{amssymb}

\usepackage[
      colorlinks=true,
      linkcolor=blue,
      urlcolor=blue,
      filecolor=black,
      citecolor=red,
      pdfstartview=FitV,
      pdftitle={},
        pdfauthor={ Michael Gutperle, John D. Miller},
        pdfsubject={},
        pdfkeywords={},
        pdfpagemode={},
        bookmarksopen=true
      ]{hyperref}

\usepackage{color}
\usepackage{graphicx}

\usepackage{sectsty}
\sectionfont{\large}

\renewcommand{\thefootnote}{\fnsymbol{footnote}}
\renewcommand{\thanks}[1]{\footnote{#1}}
\newcommand{\starttext}{
\setcounter{footnote}{0}
\renewcommand{\thefootnote}{\arabic{footnote}}}
\newcommand{\bea}{\begin{eqnarray}}
\newcommand{\eea}{\end{eqnarray}}
\newcommand{\ee}{\end{equation}}
\newcommand{\be}{\begin{equation}}

\newcommand{\no}{\nonumber}

\numberwithin{equation}{section}

\def\det{{\rm det}}

\usepackage{ dsfont } 

\renewcommand{\Im}{\operatorname{Im}}
\renewcommand{\Re}{\operatorname{Re}}

\def\no{\nonumber}

\long\def\symbolfootnote[#1]#2{\begingroup%
\def\thefootnote{\fnsymbol{footnote}}\footnote[#1]{#2}\endgroup}

\begin{document}
\setlength{\baselineskip}{18pt}

\starttext
\setcounter{footnote}{0}

\begin{flushright}
\today
\end{flushright}

\bigskip

\begin{center}

{\Large \bf   Entanglement entropy at holographic interfaces}

\vskip 0.4in

{\large   Michael Gutperle and John D. Miller}

\vskip .2in

{ \it Department of Physics and Astronomy }\\
{\it University of California, Los Angeles, CA 90095, USA}\\[0.5cm]

\bigskip
\href{mailto:gutperle@physics.ucla.edu}{\texttt{gutperle}}\texttt{, }
\href{mailto:johnmiller@physics.ucla.edu}{\texttt{johnmiller@physics.ucla.edu}}

\bigskip

\bigskip

\end{center}
 
\begin{abstract}

\setlength{\baselineskip}{18pt}

In this note we calculate the holographic entanglement entropy in the presence of a conformal interface for a  geometric configuration in which  the entangling region  ${\cal A}$   lies on one side of the interface.
 For the supersymmetric Janus solution we find exact agreement between the holographic and CFT  calculation of the entanglement entropy.

\end{abstract}

\setcounter{equation}{0}
\setcounter{footnote}{0}


\newpage

 
 \section{Introduction}
 
 In two-dimensional conformal field theories the folding trick \cite{Oshikawa:1996dj,Bachas:2001vj} allows one to map the problem of the construction of a conformal interfaces between $CFT_1$ and  $CFT_2$  to the  construction of conformal boundary state in the folded product $CFT_1\otimes \overline{CFT}_2$. This construction has been used to construct conformal interfaces for free compactified bosons and more general CFTs, see e.g. \cite{Bachas:2001vj,Fuchs:2007tx,Bachas:2007td,Quella:2006de,Gang:2008sz,Bachas:2012bj,Frohlich:2009gb, Satoh:2011an,Brunner:2013ota}. In general the entanglement entropy of an entangling region ${\cal A}$ in the presence of an interface depends on the location of the interface with respect to ${\cal A}$. For a region  of length $L$ which is placed symmetrically about the interface the entanglement entropy was calculated in \cite{Azeyanagi:2007qj}, where it was found that the logarithmically divergent term is independent of the  interface and the constant term was related to the boundary entropy or $g$ function \cite{Affleck:1991tk} of the folded boundary CFT.
 
 In this note we are interested in a different geometrical setup where the entangling region ${\cal A}$ is the half-space lying on one side of the interface.  This entanglement entropy has been calculated using the replica trick in \cite{Sakai:2008tt} for a system  corresponding to a compact boson whose compactification radius $R$  jumps across the interface. In \cite{Brehm:2015lja}  the analogous  calculation was performed for interfaces in the two-dimensional Ising model. We will review some of these results in section \ref{sec2}.

Janus solutions \cite{Bak:2003jk,D'Hoker:2006uu,Clark:2004sb,D'Hoker:2007xy} are holographic realizations of conformal interfaces\footnote{See  \cite{Karch:2000gx,DeWolfe:2001pq,Aharony:2003qf} for other approaches to describe interfaces in AdS.}. It is natural to use the Ryu-Takayanagi prescription \cite{Ryu:2006bv,Ryu:2006ef} to calculate entanglement entropy for these solutions and compare the results to the CFT calculation. For the symmetric entangling surface this was done in \cite{Azeyanagi:2007qj} using the non-supersymmetric Janus solution in three dimensions and in \cite{Chiodaroli:2010ur} using  the supersymmetric Janus solution in six dimensions which is locally asymptotic to $AdS_3\times S^3$.  In this note we calculate the holographic entanglement entropy where the entangling region ${\cal A}$ is the half space which ends at the interface.
 
 The structure of this note is as follows: In section \ref{sec2} we review the CFT  calculation of the entanglement entropy. In \ref{sec3} we calculate the entanglement entropy for the non-supersymmetric Janus solution. In section \ref{sec4} we perform the same calculation for the supersymmetric Janus solution. We present a discussion of our results and some possible avenues for future research in section \ref{sec5}. Some details of the supersymmetric solutions are delegated to appendix \ref{appa}. 
 
 \section{CFT Interfaces  and entanglement entropy}\label{sec2}
 
 In this section we review the results of \cite{Sakai:2008tt}  and \cite{Brehm:2015lja} on the CFT  calculation of the entanglement entropy in the presence of a conformal interface.
 The entanglement entropy of a region ${\cal A}$ is defined as the von Neumann entropy of the reduced density  matrix $\rho_{\cal A}=tr_{\bar{\cal A}} |0\rangle\langle 0|$. A well-known calculational method  relates the entanglement entropy to a specific limit of Renyi entropies
 \begin{align}\label{renyi}
 S_{\cal A} &= -{\partial \over \partial K} {\rm tr}  \left.\rho_{\cal A}^K \right|_{K=1}
 \end{align}

 \begin{figure}[!t]
  \centering
  \includegraphics[width=115mm]{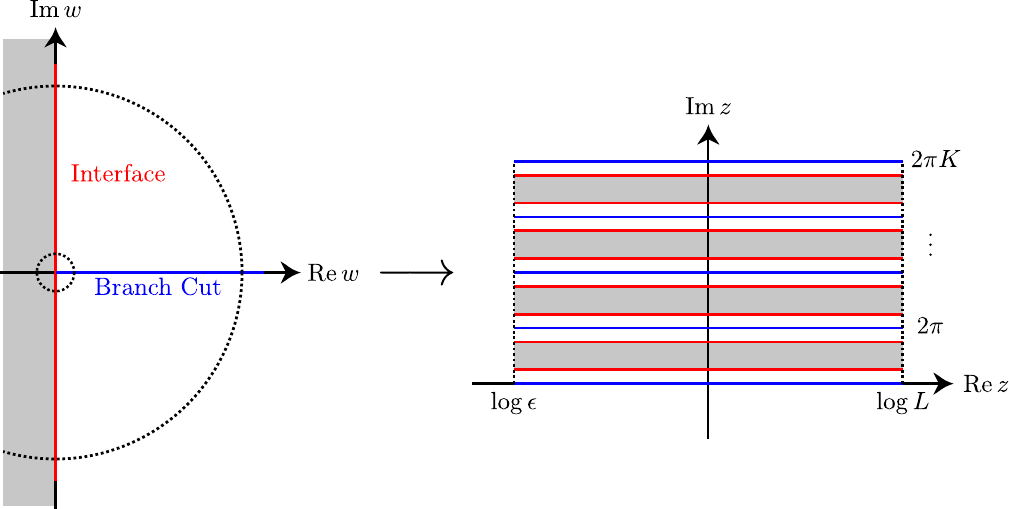}
  \caption{The  map $z=\ln w$ maps the K-sheeted Riemann surface with an interface at $\Re(w)=0$ and the branch cut at $\Im(z)=0, \Re(w)>0$  to  the geometry on the right. The dotted circles on the left correspond to the UV cutoff at $|w|=\epsilon$ and the IR cutoff at $|w|=L$. This figure was adapted from \cite{Brehm:2015lja}.}
  \label{fig:replica}
\end{figure}

 The Renyi entropies can be calculated by the replica trick in which the  trace (\ref{renyi}) is represented as a path integral over a K-sheeted Riemann surface where the branch cuts run along ${\cal A}$. The entanglement entropy can then be calculated from the partition function  $Z(K)$ on the K-sheeted Riemann surface as follows
 \begin{align}
 S_{\cal A}&= (1-\partial_K) \left.\log Z(K)\right|_{K=1}
 \end{align} 
 It has been shown in \cite{Sakai:2008tt}  that in the presence of an interface the K-sheeted partition function $Z(K)$ can be calculated by mapping the K-sheeted Riemann surface via $z=\ln w$ to a covering space  (see figure \ref{fig:replica}).
Introducing an UV  cutoff $\epsilon$ and IR cutoff $L$ and imposing periodic boundary conditions for simplicity, the K-th replica partition function  becomes
 \begin{align}\label{zkformula}
 Z(K) = Tr_{CFT_1} \Big( I_{1,2} e^{- t H_2} I_{2,1} e^{-tH_1} \Big)^K
 \end{align} 
 Where $t=2\pi^2 / \log(L/\epsilon)$ and 
 \begin{align}
 H_i&= L^{(i)}_0+\bar L^{(i)}_0-{1\over 12} c^{(i)},\quad  i=1,2
 \end{align}
 are the Hamiltonians of $CFT_1$ and $CFT_2$ respectively.  The interface operator $ I_{1,2} $ 
 maps states from $CFT_1$ to $CFT_2$ and the operator $ I_{2,1}=(I_{1,2})^\dagger$ is the conjugate interface which maps     $CFT_2$ to $CFT_1$.

 In \cite{Sakai:2008tt} the expression (\ref{zkformula}) has been determined and the entanglement entropy has been calculated for  the $c=1$ permeable 
 interface of    a compact boson  whose radius jumps from $R_1$ to $R_2$, first introduced in 
 \cite{Bachas:2001vj}.  After doubling the interface is mapped to a D1 brane  inside a 
 rectangular torus of radius $R_1$ and $R_2$ which is  winding $k_1$ times around the $R_1$ 
 cycle and $k_2$ times around the $R_2$ cycle.   The result for the entanglement entropy calculated in  \cite{Sakai:2008tt} is 
 \begin{align}\label{saboson}
 S_{\cal A} = {1\over 2} \sigma( |s| ) \log {L\over \epsilon} - \log| k_1 k_2| 
 \end{align}
  Where $ s$  is given by
 \begin{align}\label{sdefa}
 s&= \sin 2\theta_+, \quad \theta_+ = \arctan {k_2 R_2 \over k_1 R_1}
 \end{align}
 The function $\sigma(x)$  can be expressed as an integral or in terms of dilogarithm functions
 \begin{align}\label{sigmax}
 \sigma(x)&= {x\over 2} -{2\over \pi^2} \int_0^\infty  dz  \; z\Big( \sqrt{1+(x/\sinh z)^2}-1\Big)\nonumber \\
 &={1\over 6} +{x\over 3} +{1\over \pi^2}\Big( (x+1)\log(x+1) \log x+ (x-1) {\rm Li}_2(1-x)+ (x+1) {\rm Li}_2 (-x)\Big)
 \end{align}
 Note that the interface which corresponds to the Janus solution has $k_1=k_2=1$ and hence the constant term in (\ref{saboson}) vanishes. Furthermore the case of an identity defect (i.e. $R_1=R_2$) then corresponds to $\theta_+=\pi/4$ for which $\sigma(1) = {1/3}$ and formula (\ref{saboson})  agrees with the standard universal results for the entanglement entropy in a single vacuum CFT with $c=1$.  The complicated dependence of the entanglement entropy on $s$ given by the function $\sigma(|s|)$ simplifies considerably if the free boson interface is combined with a free fermion in a supersymmetric fashion as pointed out in a recent paper  \cite{Brehm:2015lja}. This is due to an extensive cancellation between bosonic and fermionic oscillators in $Z(K)$. The entanglement entropy for a supersymmetric interface in a $c=3/2$ CFT of a compact boson and a free fermion is given by \cite{Brehm:2015lja}
 \begin{align}\label{brunnera}
 S^{susy}_{\cal A} = {1\over 2} s   \log {L\over \epsilon} - \log| k_1 k_2| 
 \end{align}

 \section{Non-supersymmetric Janus solution}\label{sec3}
 
 The three-dimensional Janus solution was constructed in \cite{Bak:2007jm}. The starting point is a 
 three-dimensional gravity with negative cosmological constant coupled to a massless scalar (e.g. the dilaton field)
 \begin{align}
 S={1\over 16 \pi G_N} \int d^3 x \sqrt{g} \Big(R- \partial_\mu \phi\partial^\mu\phi+{2\over L^2}\Big)
 \end{align}
 The Janus solution solves the equations of motion coming from this action and is given by
 \begin{align}\label{nonsusymet}
 ds^2= L^2\Big( d\mu^2+ f(\mu) {dz^2-dt^2\over z^2}\Big)
 \end{align}
 where 
\begin{align}
f(\mu)= {1\over 2}\Big(1+ \sqrt{1-2\gamma^2}\cosh(2\mu)\Big)
\end{align}
and
\begin{align}
\phi(\mu)= \phi_0 -\sqrt{2}\,\tanh^{-1}\left( {-1+\sqrt{1-2\gamma^2}\over \sqrt{2}\gamma} \tanh\mu\right)
\end{align}
The solution depends on one parameter $\gamma$. The holographic solution corresponds to an interface connecting two half spaces which are reached  on the boundary of the spacetime  
 by taking $\mu\to \pm \infty$. The massless scalar $\phi$ takes two asymptotic values in this limit and as shown in \cite{Chiodaroli:2010ur} the jump  in $\phi$ can be identified with the jump in the radius of the free boson
 \begin{align}\label{sdefb}
 \frac{R_2}{R_1}=\frac{\lim_{\mu\rightarrow+\infty}e^{-\phi/2}}{\lim_{\mu\rightarrow-\infty}e^{-\phi/2}}= \exp\left\{\sqrt{2}\,\tanh^{-1} \left({-1+\sqrt{1-2\gamma^2}\over \sqrt{2}\gamma} \right)\right\}
 \end{align}

\begin{figure}[!t]
  \centering
  \includegraphics[width=105mm]{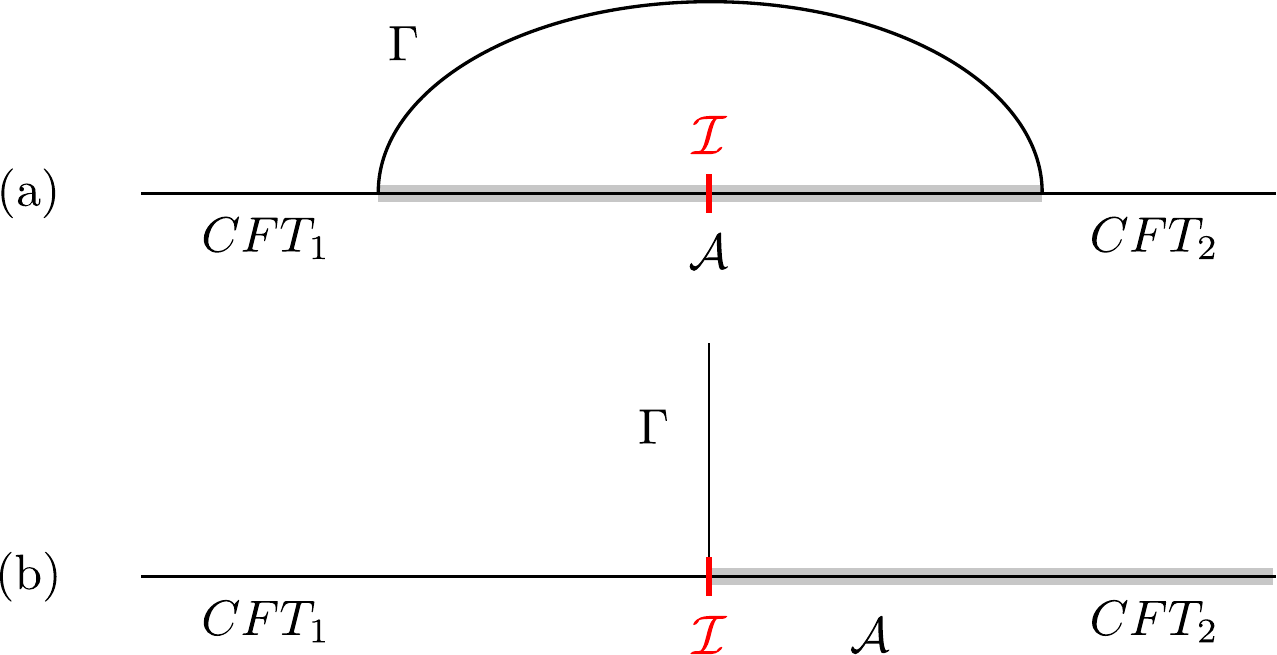}
  \caption{Two different geometries for the entangling region $\cal{A}$ and interface ${\cal I}$: (a) the entangling region is placed symmetrically about the interface, (b) the entangling surface is on one side of the interface. $\Gamma$ is a sketch of the respective RT minimal surfaces in the bulk.}
  \label{fig:entangle}
\end{figure}

According to the Ryu-Takayanagi prescription the holographic entanglement entropy  is determined by finding the area of a minimal surface (at constant time) which at the boundary of the bulk spacetime coincides with the boundary $\partial {\cal A}$ of the entangling region ${\cal A}$. In this  note we calculate the  entanglement entropy  for the entangling region on one side of the interface. We give a sketch of this geometry (b) in figure \ref{fig:entangle} and contrast it with the symmetric case depicted in (a).
 
 In three dimensions the minimal surface $\Gamma$ at $t=0$ is a curve and we have to choose an embedding. The appropriate embedding  for the case at hand turns out to be  $\mu=\mu(z)$. For this choice the induced line element leads to the following action
 \begin{align}\label{snonsusy}
A[\Gamma]=  \int dz \sqrt{ {f(\mu)\over z^2} +\left({\partial \mu\over \partial z}\right)^2}
\end{align}
The minimal area is found by solving the Euler-Lagrange equation which follows from (\ref{snonsusy})  
\begin{equation}\
f'(\mu)\left(\frac{1}{z^2}+\frac{(\partial_z\mu)^2}{f(\mu)+z^2(\partial_z\mu)^2}\right)- \frac{2}{z}\frac{f(\mu)\left(\partial_z\mu+z\partial^2_z\mu\right)}{f(\mu)+z^2(\partial_z\mu)^2}=0
\end{equation}
A simple solution  of the Euler-Lagrange equation is given by
\begin{align}
{\partial \mu\over \partial z} =0, \quad f'(\mu)=0
\end{align}
Hence $\mu$ is constant and the second equation is solved by $\mu=0$. 
It is easy to see that this solution is indeed an absolute minimum for the length, as $\mu=0$ minimizes the first term and $\partial_z\mu=0$ minimizes the second term under the square root in   the functional (\ref{snonsusy}).
The holographic entanglement entropy is then given by
\begin{align}
S_{\rm hol}&= {L\over 4 G_N}\sqrt{f(0)}\int {dz\over z}\nonumber \\
&= {c\over 6\sqrt{2}} \sqrt{1+ \sqrt{1-2\gamma^2}}\, \log {L\over \epsilon}
\end{align}
where we have regulated the divergent integral over $z$ and used  $c= {3L\over 2 G_N}$. In order to compare the functional dependence it is useful to expand the result as a power series in terms of small $\gamma$, for the holographic entanglement entropy one finds
\begin{align}
S_{\rm hol} =\Big( 1 - {1\over 4} \gamma^2 -{5\over 32} \gamma^4 +o(\gamma^6)\Big)  \;  {c\over 6} \log {L\over \epsilon}
\end{align}
We can compare this to the CFT result for the entanglement entropy (\ref{saboson}). We set $k_1=k_2=1$ which makes the constant term  vanish, expanding (\ref{sigmax}) around $s=1$ gives
\begin{align}\label{ours}
\frac{1}{2}\sigma(s) &= \frac{1}{6}-\frac{1}{8}(1-s)-\frac{1}{4\pi^2}(1-s)^2+o[(1-s)^3]\nonumber\\
&= \frac{1}{6}-\frac{1}{16}\gamma^2-\left(\frac{11}{192}+\frac{1}{16\pi^2}\right)\gamma^4+o[\gamma^6]
\end{align}
where we have used the expansion
\begin{align}
s= 1-{\gamma^2\over 2}-{11\over 24}\gamma^4+o(\gamma^6)
\end{align}
which follows from (\ref{sdefa}) and (\ref{sdefb}). Using this expansion in the CFT entanglement entropy (\ref{saboson}) and restoring a general value for the central charge (i.e. by considering $c$ copies of the single boson) gives
\begin{equation}\label{satoh}
S_{\rm CFT} = \Big( 1-{3\over 8}\gamma^2 -\left( {11\over 32}+ {3\over 8\pi^2} \right) \gamma^4 + o(\gamma^6)\Big) \; {c\over 6} \log {L\over \epsilon}
\end{equation}
Comparing (\ref{ours}) and (\ref{satoh}) shows that the two expressions only agree for $\gamma=0$ which corresponds to the case where no interface is present. This result is to be contrasted with  result \cite{Azeyanagi:2007qj} for the symmetric entangling region where agreement of the CFT and the holographic  entanglement entropy up to order $\gamma^2$ was found.

 \section{Supersymmetric Janus solution}\label{sec4}
 
The supersymmetric Janus solution of type IIB which is locally asymptotic to $AdS_3\times S^3 \times M_4$   was constructed in   \cite{Chiodaroli:2009yw} (see \cite{Kumar:2002wc,Kumar:2003xi} for some earlier work in this direction and \cite{Chiodaroli:2009xh,Chiodaroli:2011nr} for generalizations). Some aspects of the solutions are reviewed in appendix \ref{appa} for the convenience of the reader.
The metric for the solution takes the following form
\begin{align}
ds^2& = f_1^2 {dz^2-dt^2\over z^2}+ f_2^2 (d\phi_1^2+\sin^2\phi_1 d\phi_2) +  f_3^2 ds_{M_4}^2+ \rho^2(dx^2+dy^2)
\end{align}
We parametrize the minimal surface for the entanglement entropy by $t=0$ and $x=x(z,y)$, i.e. the eight-dimensional surface is spanned by $\xi^a=\{z,y, \phi_1,\phi_2\}$ and the four coordinates of $M_4$.  The induced metric is then given by
\begin{align}
\gamma_{ab} = {\partial x^\mu\over \partial \xi^a} {\partial x^\nu\over \partial \xi^b} g_{\mu\nu}
\end{align}
and the action for the minimal surface is 
\begin{align}\label{Ssusy}
S&= \int d^8 \xi \sqrt{\det \gamma}\\
&=  \int_{M_4}  dV \int d\phi_1 \;d\phi_2 \,\sin\phi_1 \int dz \;dy \,{1\over z} f_2^2 f_3^4 \rho \sqrt{f_1^2\left(1+(\partial_yx)^2\right)+z^2 \rho^2 (\partial_z x)^2}
\end{align}
The Euler-Lagrange equation following from (\ref{Ssusy}) is given by

\begin{align}\label{elsusy}
0=&\quad {1\over z} \partial_x \left( f_2^2 f_3^4 \rho \sqrt{f_1^2\left(1+(\partial_yx)^2\right)+z^2 \rho^2 (\partial_z x)^2}\right)\nonumber\\
&-\partial_z\left( { f_2^2 f_3^4 \rho^3  z  \partial_z x\over \sqrt{f_1^2\left(1+(\partial_yx)^2\right)+z^2 \rho^2 (\partial_z x)^2}}\right) 
-\partial_y\left( { f_2^2 f_3^4 \rho^2  f_1^2  \partial_y x\over  z \sqrt{f_1^2\left(1+(\partial_yx)^2\right)+z^2 \rho^2 (\partial_z x)^2}}\right)
\end{align}
While it seems formidable to find a solution to (\ref{elsusy}) a simple solution can be found by setting
\begin{align}
x(z,y)=x_0
\end{align}
for which it is straightforward to verify that (\ref{elsusy}) reduces to
\begin{align}\label{eqofmb}
\partial_x \log\Big(f_1^2 f_2^4 f_3^8 \rho^2\Big)|_{x=x_0} =0
\end{align}
which has to be valid for all values of $y$.  Plugging in the solution for the metric factors found in appendix \ref{appa}
one finds
\begin{align}
f_1^2 f_2^4 f_3^8 \rho^2= 16L^4 \cosh^2\psi \cosh^2 \theta \cosh^2(x+\psi) \sin^4 y
\end{align}
and hence (\ref{eqofmb}) is satisfied if $x_0=-\psi$. Since the expression under the square root in the action functional (\ref{Ssusy}) is the sum of positive terms which are all minimized by the  solution, we have indeed an absolute minimum as demanded by the Ryu-Takayanagi prescription.
 For the solution the area is given by
\begin{align}
A&=   \int_{M_4}  dV \,Vol(S_2) \int dy \sin^2 y \int {dz\over z}  \,4 L^2 \cosh \psi \cosh\theta \nonumber\\
&=  Vol(S_3) Vol(M_4) 4 L^2  \cosh \psi \cosh\theta  \log{L\over \epsilon}\nonumber\\
&= 8\pi^2 L^2 Vol(M_4) \cosh\psi\cosh\theta\, \log{L\over \epsilon}
\end{align}
As reviewed in appendix \ref{appa} the central charge  of the dual CFT is given in terms of the parameters 
\begin{align}
c&= {3 \times 32 \pi^3  Vol(M_4)  L^2 \over \kappa_{10}^2} \cosh^2 \psi \sinh^2\theta
\end{align}
Using the result of the area the holographic entanglement entropy can then be expressed as
\begin{align}\label{holosusy}
S_A&= {A\over 4 G_N^{(10)}}\nonumber\\
&={1\over \cosh\theta \cosh\psi} \;  {c\over 6}  \log{L\over \epsilon} 
\end{align}
where we used the identification ${1/ 16 G_N^{(10)} }={ 1/2 \kappa_{10}^2}$. In order to compare the holographic result (\ref{holosusy})  to  the CFT (\ref{brunnera}) we have to set  $\theta=0$
which on the CFT side corresponds to an interface where only the radius of $M_4$ jumps and 
there is no jump of the RR modulus \cite{Chiodaroli:2010ur}. The jump of the radius can be identified with the parameter $\psi$ of the supergravity solution as follows  \cite{Chiodaroli:2010ur}
\begin{align}
{R_2\over R_1} &= e^{\psi}
\end{align}
 and hence
 \begin{align}
 2\cosh \psi &= {r_+\over r_-} +{r_-\over r_+} 
 \end{align}
 The identification of $s$ is given by
 \begin{align}
 s&= \sin 2\theta_+ = {2 r_+ r_-\over r_+^2+r_-^2} = {1\over \cosh \psi}
 \end{align}
 Hence in this special case the  holographic entanglement entropy (\ref{holosusy}) becomes
 \begin{align}
 S_A& = {c\over 6} s \log{L\over \epsilon} 
 \end{align}
  which is in exact  agreement with the CFT result (\ref{brunnera})
  if  we replace  the value   $c=3/2$  for a real boson and a real fermion with the general value of the central charge. As far as this identification is concerned in our case  the symmetric orbifold CFT which is dual to supergravity on $AdS_3\times S_3\times M_4$ can simply be viewed as $4N=4 Q_5 Q_1$ copies of the $c=3/2$ system.
 
 \section{Discussion}\label{sec5}
 
 In this note the holographic entanglement entropy was calculated for a surface ${\cal A}$ which lies on one side of a conformal interface.   It is interesting to contrast the result (\ref{saboson}) with the result for the entanglement entropy for a surface which is lying symmetrically across the interface:
  \begin{align}\label{sentsymm}
 S^{\rm sym}_{\cal A}=  {c\over 6} \log {L\over \epsilon}+ \ln g_B 
 \end{align}
  Note that for the geometric setup discussed in this note the logarithmically divergent term does not have a universal prefactor $c/6$ but depends on the parameters of the interface via the function $\sigma(|s|)$.  This difference makes sense as the interface is located at the boundary between ${\cal A}$ and its complement, where the entanglement between the two regions is strongest.
  
It is also interesting to compare the holographic calculations of  the entanglement entropy for the two cases.  In  \cite{Azeyanagi:2007qj}  the non-supersymmetric Janus solution was used to calculate (\ref{sentsymm}) and in particular the holographic boundary entropy $\ln g_B$ was calculated. A comparison with the CFT calculation led to an agreement of $\ln g_B$ to first nontrivial order in the deformation parameter $\gamma$.  In section \ref{sec3} we found that in our case the result disagrees even to the lowest nontrivial order in $\gamma$.
 
This state is to be contrasted with the supersymmetric Janus solution where both for the symmetric entangling region \cite{Chiodaroli:2010ur} and the one-sided case calculated in section \ref{sec4} the CFT  and the holographic entanglement entropy agree.  Note that the CFT and the gravity calculations are performed at very different points in the moduli space of the dual  CFT. It is likely that the high degree of supersymmetry allows the extrapolation of  the results from one point  to the other\footnote{In a recent paper \cite{Erdmenger:2015spo} the entanglement entropy in a (nonsupersymmetric) holographic model of the Kondo model was calculated and agreement with field theory results was found.}.

The supersymmetric Janus solution depends on two parameters $\theta$ and $\psi$ and we set $\theta=0$ for the comparison. The parameter $\theta$ corresponds to an RR modulus and consequently to a twist field in the symmetric orbifold CFT. It would be interesting to see whether the CFT calculation can be performed for a general interface operator $I_{12}$ which includes a jump in the twist field.   

Recently the CFT at the symmetric orbifold point has been conjectured to be dual to a higher spin theory 
\cite{Gaberdiel:2014cha,Gaberdiel:2015mra}. The region in moduli space where supergravity is valid is far removed from this point. Supersymmetry seems to make the result of the entanglement entropy independent of where on its moduli space  the theory is. It would be interesting to investigate whether it is possible to construct  the relevant interface theories in the Chern-Simons formulation following  \cite{Gutperle:2013ema} and calculate the entanglement entropy following the proposals relating the entanglement entropy and the Wilson loop in higher spin theory \cite{Ammon:2013hba,deBoer:2013vca,deBoer:2014sna}.

 \section*{Acknowledgements}
  This work was supported in part by National Science Foundation grant PHY-13-13986. The work of MG  was in part supported by a fellowship of the Simons Foundation. MG thanks the Institute for Theoretical Physics, ETH Z\" urich, for hospitality while part of this work was performed.

 \newpage

\appendix

\section{Supersymmetric Janus solution}\label{appa}
 In this appendix we review the details of the supersymmetric Janus solution for the convenience of the reader.  This solution  was first constructed in \cite{Chiodaroli:2009yw} and generalized in \cite{Chiodaroli:2009xh,Chiodaroli:2011nr}, where more details can be found.
 The ten-dimensional Janus  metric is constructed as a fibration of $AdS_2 \times S^2 \times M_4$, where $M_4$ is either $T_4$ or $K_3$, over a two dimensional Riemann surface $\Sigma$
\bea\label{bpsmeta}
ds^{2} &=& f_{1}^{2 } ds^{2}_{AdS_{2}} + f^{2}_{2}ds^{2}_{S^{2}} + f^{2}_{3}ds^{2}_{M_4}  + \rho^{2 }dz  d\bar z
\eea
All fields
depend on the coordinates $z,\bar z$ of  the surface  $\Sigma$.  For the supersymmetric
Janus solution we choose $\Sigma$ as an infinite strip as follows
\bea
w=x+i y, \quad \quad  x\in {}[-\infty,+\infty{} ], \;\; y\in {}[0,\pi{}]
\eea
The boundaries of the strip  are located at $y=0,\pi$.
The supersymmetric Janus solution depends on four parameters $k,L,\theta$ and $\psi$.
The dilaton and axion are given, respectively, by
\bea
 e^{-2\phi} &=&  k^4 {\cosh^2 (x + \psi)  {\rm sech}^2 \psi + \big(  \cosh^2 \theta -  {\rm sech}^2  \psi \big) \sin^2 y \over \big( \cosh  x  - \cos y \tanh \theta \big)^2} \\
\chi &=& - {k^2 \over 2}  {  \sinh 2 \theta  \sinh x - 2 \tanh \psi \cos y \over \cosh x \cosh \theta - \cos y \sinh \theta} \eea
The metric factors on $\Sigma$ and $M_4$ are  
\bea
\rho^4&=& { e^{- \phi} } {L^2\over k^2}  { \cosh^2 x \cosh^2 \theta - \cos^2 y \sinh^2 \theta \over \cosh^2 (x + \psi) } \cosh^4 \psi \no\\
f_3^4 &=&e^{- \phi} {4\over k^2} {\cosh x \cosh\theta-\cos y \sinh \theta \over \cosh x \cosh\theta+\cos y \sinh \theta  }
\eea
The following expressions for the  $AdS_2$ and $S^2$ metric factors will be useful,
\bea
{f^2_1 \over \rho^2} &=& {\cosh^2 \big(x + \psi \big)\over \cosh^2 \theta \cosh^2 \psi } \no \\
{\rho^2\over f_2^2}  &=&{1\over \sin^2 y} + {\cosh^2 \theta \cosh^2 \psi - 1 \over  \cosh^2 \big(x + \psi \big) } \label{relmetr3}
\eea
While the form of the antisymmetric tensor fields is not essential, we quote the expressions for the $D1$ and $D5$ brane charges from \cite{Chiodaroli:2009yw}.
\bea
 Q_{D5}&=& 4 \pi^2 k L Vol(M_4)  \cosh \psi \;  \cosh \theta \no\\
 Q_{D1}&=& {16 \pi^2 L\over k} \cosh \psi \;  \cosh \theta
 \eea
The  dual CFT is a  $\mathcal{N}=(4,4)$ SCFT  which, at a particular point of its moduli space,
is a  $(M^4)^{Q_{D1} Q_{D5}} / S_{Q_{D1} Q_{D5}}$ orbifold.
 The central charge $c$  of this  CFT takes the following form
\bea\label{centralc}
c&=& {6\over 4 \pi  k_{10}^2}  Q_{D1}Q_{D5}  = {3 \times 32 \; \pi^3  Vol(M_4)\; L^2 \over k_{10}^2} \cosh^2\psi\;  \cosh^2\theta \eea


\newpage 
 
 
\begin{thebibliography}{99}




\bibitem{Oshikawa:1996dj}
  M.~Oshikawa and I.~Affleck,
  ``Boundary conformal field theory approach to the critical two-dimensional Ising model with a defect line,''
  Nucl.\ Phys.\ B {\bf 495} (1997) 533
  [cond-mat/9612187].



\bibitem{Bachas:2001vj}
  C.~Bachas, J.~de Boer, R.~Dijkgraaf and H.~Ooguri,
  ``Permeable conformal walls and holography,''
  JHEP {\bf 0206} (2002) 027
  [hep-th/0111210].


\bibitem{Fuchs:2007tx}
  J.~Fuchs, M.~R.~Gaberdiel, I.~Runkel and C.~Schweigert,
  ``Topological defects for the free boson CFT,''
  J.\ Phys.\ A {\bf 40} (2007) 11403
  [arXiv:0705.3129 [hep-th]].


\bibitem{Bachas:2007td}
  C.~Bachas and I.~Brunner,
  ``Fusion of conformal interfaces,''
  JHEP {\bf 0802} (2008) 085
  [arXiv:0712.0076 [hep-th]].

\bibitem{Quella:2006de}
  T.~Quella, I.~Runkel and G.~M.~T.~Watts,
  ``Reflection and transmission for conformal defects,''
  JHEP {\bf 0704} (2007) 095
  [hep-th/0611296].



\bibitem{Gang:2008sz}
  D.~Gang and S.~Yamaguchi,
  ``Superconformal defects in the tricritical Ising model,''
  JHEP {\bf 0812} (2008) 076
  [arXiv:0809.0175 [hep-th]].


  
\bibitem{Bachas:2012bj}
  C.~Bachas, I.~Brunner and D.~Roggenkamp,
  ``A worldsheet extension of O(d,d:Z),''
  JHEP {\bf 1210} (2012) 039
  [arXiv:1205.4647 [hep-th]].
 

\bibitem{Frohlich:2009gb}
  J.~Frohlich, J.~Fuchs, I.~Runkel and C.~Schweigert,
  ``Defect lines, dualities, and generalised orbifolds,''
  arXiv:0909.5013 [math-ph].
 

 
\bibitem{Satoh:2011an}
  Y.~Satoh,
  ``On supersymmetric interfaces for string theory,''
  JHEP {\bf 1203} (2012) 072
  [arXiv:1112.5935 [hep-th]].
  
 
    
  
\bibitem{Brunner:2013ota}
  I.~Brunner, N.~Carqueville and D.~Plencner,
  ``Orbifolds and topological defects,''
  Commun.\ Math.\ Phys.\  {\bf 332} (2014) 669
  [arXiv:1307.3141 [hep-th]].
  
 
 
\bibitem{Azeyanagi:2007qj}
  T.~Azeyanagi, A.~Karch, T.~Takayanagi and E.~G.~Thompson,
  ``Holographic calculation of boundary entropy,''
  JHEP {\bf 0803} (2008) 054
  [arXiv:0712.1850 [hep-th]].

 
 
\bibitem{Affleck:1991tk}
  I.~Affleck and A.~W.~W.~Ludwig,
  ``Universal noninteger 'ground state degeneracy' in critical quantum systems,''
  Phys.\ Rev.\ Lett.\  {\bf 67} (1991) 161.
 
 
\bibitem{Sakai:2008tt}
  K.~Sakai and Y.~Satoh,
  ``Entanglement through conformal interfaces,''
  JHEP {\bf 0812} (2008) 001
  [arXiv:0809.4548 [hep-th]].
  


\bibitem{Brehm:2015lja}
  E.~M.~Brehm and I.~Brunner,
  ``Entanglement entropy through conformal interfaces in the 2D Ising model,''
  JHEP {\bf 1509} (2015) 080
  [arXiv:1505.02647 [hep-th]].


\bibitem{Brehm:2015lja}
  E.~M.~Brehm and I.~Brunner,
  ``Entanglement entropy through conformal interfaces in the 2D Ising model,''
  JHEP {\bf 1509} (2015) 080
  [arXiv:1505.02647 [hep-th]].



\bibitem{Bak:2003jk}
  D.~Bak, M.~Gutperle and S.~Hirano,
  ``A Dilatonic deformation of AdS(5) and its field theory dual,''
  JHEP {\bf 0305} (2003) 072
  [hep-th/0304129].



\bibitem{D'Hoker:2006uu}
  E.~D'Hoker, J.~Estes and M.~Gutperle,
  ``Ten-dimensional supersymmetric Janus solutions,''
  Nucl.\ Phys.\ B {\bf 757} (2006) 79
  [hep-th/0603012].

\bibitem{Clark:2004sb}
  A.~B.~Clark, D.~Z.~Freedman, A.~Karch and M.~Schnabl,
  ``Dual of the Janus solution: An interface conformal field theory,''
  Phys.\ Rev.\ D {\bf 71} (2005) 066003
  [hep-th/0407073].



\bibitem{D'Hoker:2007xy}
  E.~D'Hoker, J.~Estes and M.~Gutperle,
  ``Exact half-BPS Type IIB interface solutions. I. Local solution and supersymmetric Janus,''
  JHEP {\bf 0706} (2007) 021
  [arXiv:0705.0022 [hep-th]].




\bibitem{Karch:2000gx}
  A.~Karch and L.~Randall,
  ``Open and closed string interpretation of SUSY CFT's on branes with boundaries,''
  JHEP {\bf 0106} (2001) 063
  [hep-th/0105132].

\bibitem{DeWolfe:2001pq}
  O.~DeWolfe, D.~Z.~Freedman and H.~Ooguri,
  ``Holography and defect conformal field theories,''
  Phys.\ Rev.\ D {\bf 66} (2002) 025009
  [hep-th/0111135].

\bibitem{Aharony:2003qf}
  O.~Aharony, O.~DeWolfe, D.~Z.~Freedman and A.~Karch,
  ``Defect conformal field theory and locally localized gravity,''
  JHEP {\bf 0307} (2003) 030
  [hep-th/0303249].


\bibitem{Ryu:2006bv}
  S.~Ryu and T.~Takayanagi,
  ``Holographic derivation of entanglement entropy from AdS/CFT,''
  Phys.\ Rev.\ Lett.\  {\bf 96} (2006) 181602
  [arXiv:hep-th/0603001].


\bibitem{Ryu:2006ef}
  S.~Ryu and T.~Takayanagi,
  ``Aspects of holographic entanglement entropy,''
  JHEP {\bf 0608} (2006) 045
  [arXiv:hep-th/0605073].



\bibitem{Chiodaroli:2010ur}
  M.~Chiodaroli, M.~Gutperle and L.~Y.~Hung,
  ``Boundary entropy of supersymmetric Janus solutions,''
  JHEP {\bf 1009} (2010) 082
  [arXiv:1005.4433 [hep-th]].

\bibitem{Bak:2007jm}
  D.~Bak, M.~Gutperle and S.~Hirano,
  ``Three dimensional Janus and time-dependent black holes,''
  JHEP {\bf 0702} (2007) 068
  [hep-th/0701108].


\bibitem{Chiodaroli:2009yw}
  M.~Chiodaroli, M.~Gutperle and D.~Krym,
  ``Half-BPS Solutions locally asymptotic to AdS(3) x S**3 and interface conformal field theories,''
  JHEP {\bf 1002} (2010) 066
  [arXiv:0910.0466 [hep-th]].



\bibitem{Kumar:2002wc}
  J.~Kumar and A.~Rajaraman,
  ``New supergravity solutions for branes in  $AdS_3 \times  S^3$,''
  Phys.\ Rev.\  D {\bf 67} (2003) 125005
  [arXiv:hep-th/0212145].

\bibitem{Kumar:2003xi}
  J.~Kumar and A.~Rajaraman,
  ``Supergravity solutions for  $AdS_3\times   S^3$ branes,''
  Phys.\ Rev.\  D {\bf 69} (2004) 105023
  [arXiv:hep-th/0310056].


\bibitem{Chiodaroli:2009xh}
  M.~Chiodaroli, E.~D'Hoker and M.~Gutperle,
  ``Open Worldsheets for Holographic Interfaces,''
  JHEP {\bf 1003} (2010) 060
  [arXiv:0912.4679 [hep-th]].

\bibitem{Chiodaroli:2011nr}
  M.~Chiodaroli, E.~D'Hoker, Y.~Guo and M.~Gutperle,
  ``Exact half-BPS string-junction solutions in six-dimensional supergravity,''
  JHEP {\bf 1112} (2011) 086
  [arXiv:1107.1722 [hep-th]].

\bibitem{Erdmenger:2015spo} 
  J.~Erdmenger, M.~Flory, C.~Hoyos, M.~N.~Newrzella and J.~M.~S.~Wu,
  ``Entanglement Entropy in a Holographic Kondo Model,''
  arXiv:1511.03666 [hep-th].



  
\bibitem{Gaberdiel:2014cha}
  M.~R.~Gaberdiel and R.~Gopakumar,
  ``Higher Spins and  Strings,''
  JHEP {\bf 1411} (2014) 044
  [arXiv:1406.6103 [hep-th]].
  
\bibitem{Gaberdiel:2015mra}
  M.~R.~Gaberdiel and R.~Gopakumar,
  ``Stringy Symmetries and the Higher Spin Square,''
  J.\ Phys.\ A {\bf 48} (2015) 18,  185402
  [arXiv:1501.07236 [hep-th]].


\bibitem{Gutperle:2013ema}
  M.~Gutperle,
  ``A note on interface solutions in higher-spin gravity,''
  JHEP {\bf 1307} (2013) 091
  [arXiv:1302.3653 [hep-th]].



\bibitem{Ammon:2013hba}
  M.~Ammon, A.~Castro and N.~Iqbal,
  ``Wilson Lines and Entanglement Entropy in Higher Spin Gravity,''
  JHEP {\bf 1310} (2013) 110
  [arXiv:1306.4338 [hep-th]].

\bibitem{deBoer:2013vca}
  J.~de Boer and J.~I.~Jottar,
  ``Entanglement Entropy and Higher Spin Holography in AdS$_3$,''
  JHEP {\bf 1404} (2014) 089
  [arXiv:1306.4347 [hep-th]].

\bibitem{deBoer:2014sna}
  J.~de Boer, A.~Castro, E.~Hijano, J.~I.~Jottar and P.~Kraus,
  ``Higher spin entanglement and $ {\mathcal{W}}_{\mathrm{N}} $ conformal blocks,''
  JHEP {\bf 1507} (2015) 168
  [arXiv:1412.7520 [hep-th]].
  
  
  
  \end{thebibliography}
\end{document}